\documentclass[10pt]{emulateapj}
\usepackage{apjfonts}

\usepackage{amssymb,amsmath}
\usepackage{color}

\newcommand{\degree}{\ensuremath{^\circ}}

\newcommand{\chisq}{\ensuremath{\chi^2}}
\newcommand{\msun}{\ensuremath{\rm{M}_\odot}}
\newcommand{\rsun}{\ensuremath{\rm{R}_\odot}}
\newcommand{\Lsun}{\ensuremath{\rm{L}_\odot}}

\newcommand{\pastmax}{1840}

\newcommand{\BmH}{\ensuremath{2.3 \pm 0.1}} 
\newcommand{\BmHmax}{\ensuremath{-0.85 \pm 0.04}} 
\newcommand{\BmHfact}{\ensuremath{20}}

\newcommand{\MNifit}{\ensuremath{ 0.29^{+ 0.03 }_{- 0.03 } }}

\newcommand{\CotoNifit}{\ensuremath{ -1.59^{+ 0.06 }_{- 0.07 } }}

\newcommand{\FetoCofit}{\ensuremath{ -1.1^{+ 0.2 }_{- 0.4 } }} 
\newcommand{\FetoCofitupper}{\ensuremath{0.22}} 
\newcommand{\Compfitupper}{\ensuremath{3.6 \times 10^{35}}} 
\newcommand{\CompfitupperLsun}{\ensuremath{90}} 
\newcommand{\echofitupper}{\ensuremath{4.3 \times 10^{35}}} 
\newcommand{\echofitupperLsun}{\ensuremath{110}} 
\newcommand{\echofitupperPer}{\ensuremath{27}}

\newcommand{\bolfive}{\ensuremath{420 \pm 20}}

\newcommand{\VMChiSq}{\ensuremath{31.4}} 
\newcommand{\DDChiSq}{\ensuremath{302.4}} 
\newcommand{\fitallChiSq}{\ensuremath{5.4}} 
\newcommand{\noFeChiSq}{\ensuremath{7.3}} 
\newcommand{\noCoChiSq}{\ensuremath{216.5}} 
\newcommand{\echoChiSq}{\ensuremath{237.3}} 
\newcommand{\compChiSq}{\ensuremath{427.5}}

\newcommand{\VMdof}{\ensuremath{4}} 
\newcommand{\DDdof}{\ensuremath{4}} 
\newcommand{\fitalldof}{\ensuremath{2}} 
\newcommand{\noFedof}{\ensuremath{3}} 
 
\newcommand{\echodof}{\ensuremath{3}} 
\newcommand{\compdof}{\ensuremath{3}}

\newcommand{\CotoNibestfitval}{\ensuremath{0.03}} 
\newcommand{\FetoCobestfitval}{\ensuremath{0.07}} 
\newcommand{\bolwavedw}{\ensuremath{4000}} 
\newcommand{\bolwaveup}{\ensuremath{17000}} 
\newcommand{\MNiassume}{\ensuremath{0.48 \pm 0.05}} 
\newcommand{\BolinOptNIR}{\ensuremath{0.60 \pm 0.08}} 
\newcommand{\stopit}{\ensuremath{0.001}} 
\newcommand{\MCoEst}{\ensuremath{ 0.012^{+ 0.002 }_{- 0.002 } }}

\newcommand{\CotoNifitopt}{\ensuremath{ -1.53^{+ 0.06 }_{- 0.06 } }}

\newcommand{\FetoCofitopt}{\ensuremath{ -1.6^{+ 0.4 }_{- 0.6 } }} 
\newcommand{\FetoCofitupperopt}{\ensuremath{0.1}}

\newcommand{\bolwavedwopt}{\ensuremath{4000}} 
\newcommand{\bolwaveupopt}{\ensuremath{9500}} 
 
\newcommand{\Bolinopt}{\ensuremath{0.39 \pm 0.05}}

\newcommand{\Blimitdate}{\ensuremath{2500}} 
\newcommand{\Vlimitdate}{\ensuremath{2500}} 
\newcommand{\RIlimitdate}{\ensuremath{2400}} 
\newcommand{\Blimit}{\ensuremath{28.9}} 
\newcommand{\Vlimit}{\ensuremath{29.1}} 
\newcommand{\RIlimit}{\ensuremath{28.6}} 

\newcommand{\varwavedw}{\ensuremath{3000}} 
\newcommand{\varwaveup}{\ensuremath{8000}} 
\newcommand{\varUlfl}{\ensuremath{6.1 \times 10^{36}}} 
\newcommand{\varBlfl}{\ensuremath{2.5 \times 10^{36}}} 
\newcommand{\varVlfl}{\ensuremath{6.1 \times 10^{36}}} 
\newcommand{\varRlfl}{\ensuremath{2 \times 10^{36}}} 
\newcommand{\varbol}{\ensuremath{1.2 \times 10^{37}}} 
\newcommand{\varUbolLsun}{\ensuremath{2000}} 
\newcommand{\varBbolLsun}{\ensuremath{1000}} 
\newcommand{\varVbolLsun}{\ensuremath{2000}} 
\newcommand{\varRbolLsun}{\ensuremath{1000}} 
\newcommand{\varbolLsun}{\ensuremath{3000}}

\begin{document}

\title{Whimper of a Bang: Documenting the Final Days of the Nearby Type Ia Supernova 2011fe}
\shorttitle{Late-time Observations of SN~2011fe}
\shortauthors{Shappee et al.}	

\author{
{B.~J.~Shappee}\altaffilmark{1,2}, 
{K.~Z.~Stanek}\altaffilmark{3,4},
{C.~S.~Kochanek}\altaffilmark{3,4},
and
{P. M.~Garnavich}\altaffilmark{5}
}

\email{bshappee@obs.carnegiescience.edu}

\altaffiltext{1}{The Observatories of the Carnegie Institution for Science, 813 Santa Barbara St., Pasadena, CA 91101, USA}
\altaffiltext{2}{Hubble, Carnegie-Princeton Fellow}
\altaffiltext{3}{Department of Astronomy, The Ohio State University, 140 West 18th Avenue, Columbus, OH 43210, USA}
\altaffiltext{4}{Center for Cosmology and AstroParticle Physics (CCAPP), The Ohio State University, 191 W.\ Woodruff Ave., Columbus, OH 43210, USA}
\altaffiltext{5}{Department of Physics, University of Notre Dame, Notre Dame, IN 46556, USA}

\begin{abstract}

Using the Hubble Space Telescope (HST) and the Large Binocular Telescope, we followed the evolution of the Type Ia supernova (SN Ia) 2011fe for an unprecedented $\pastmax$ days past $B$-band maximum light and over a factor of $7$ million in flux.  At $\pastmax$ days, the $\bolwavedw$ -- $\bolwaveup$ \AA{} quasi-bolometric luminosity is just (\bolfive{}) \Lsun{}.  
By measuring the late-time quasi-bolometric light curve, we present the first confident detection of $^{57}$Co decay in a SN Ia light curve and estimate a mass ratio of $\log (^{57}\mathrm{Co}/^{56}\mathrm{Co}) = \CotoNifit$.  We do not have a clean detection of $^{55}\mathrm{Fe}$, but find a limit of $^{55}\mathrm{Fe}/^{57}\mathrm{Co} < \FetoCofitupper$ with 99\% confidence.  These abundance ratios provide unique constraints on the progenitor system because the central density of the exploding white dwarf(s) dictates these nucleosynthetic yields.  The observed ratios strongly prefer the lower central densities of 
double-degenerate models ($^{55}\mathrm{Fe}/^{57}\mathrm{Co} = 0.27$) over the higher central densities of near-Chandrasekhar-mass single-degenerate models ($^{55}\mathrm{Fe}/^{57}\mathrm{Co} = 0.68$).  However, additional theoretical studies predicting isotopic yields from a broader range of progenitor systems are motivated by these unique observations. We will continue to observe SN 2011fe for another $\sim 600$ days with HST and possibly beyond.

\end{abstract}

\section{Introduction}
\label{sec:Intro}

Although it is generally accepted that SNe Ia are produced in close binary systems where the constituents interact, there are still two competing classes of models.  The double-degenerate (DD) model consists of a binary system with two white dwarfs (WDs) that merge either due to the removal of energy and angular momentum from the system by gravitational radiation \citep{tutukov79, iben84, webbink84} or due to the perturbations of a third (e.g., \citealp{thompson11, katz12, shappee13c, antognini14}) or fourth \citep{pejcha13} body.  The single-degenerate (SD) model consists of a WD primary and a non-degenerate secondary \citep{whelan73, nomoto82}.  Most current work on the SD model focuses on systems in which the primary accretes matter from the secondary until the primary becomes unstable to runaway nuclear burning.  

The search for observational signatures capable of distinguishing between these two models has been difficult.  Most DD models effectively have no direct signatures.  The SD model is thought to have observable consequences and some classes of SD models have been observationally ruled out as the dominant channel (e.g., \citealp{leonard07, bianco11, chomiuk16, maguire16}), but there are a number of systems that are considered to be candidate progenitor systems (e.g. U Sco and V445 Pup; \citealp{li11}) and there might be multiple channels for producing normal SNe Ia (e.g., \citealp{maguire13, yamaguchi15}). Current explosion simulations of both SD (e.g. \citealp{kasen09}) and DD (e.g. \citealp{pakmor12}) progenitors can match the observable signatures of SNe Ia around $B$-band maximum light ($t_{B {\rm max}}$).
Thus, the nature of the progenitor systems of SNe Ia remains a largely unsolved problem in modern astronomy.  

The very-late-time light curves of SNe Ia are a powerful test of these two scenarios.  The near-Chandrasekhar-mass WDs of the SD scenario have much higher central densities than the lower-mass WDs of the DD scenario, which leads to significant differences in the nucleosynthetic yields \citep{roepke12}.  Most important for this study, the higher-density near-Chandrasekhar-mass WDs produce significantly more $^{55}\mathrm{Co}$ and $^{55}\mathrm{Fe}$. 
For example, \citet{roepke12} found in their 3D simulations that the SD (DD) model produced $0.019\msun{}$  $(0.015 \msun{})$ of $^{57}$Ni and $^{57}$Co combined, and $0.013 \msun{}$ $(0.004 \msun{})$ of $^{55}$Co and $^{55}$Fe combined, respectively. 

At very-late times ($> 1050$ days after $t_{B {\rm max}}$), $^{57}\mathrm{Co}$ and $^{55}\mathrm{Fe}$ become the dominant power source for SNe Ia light curves through the following decay chains \citep{seitenzahl09},  
\begin{eqnarray}
& &^{56}\mathrm{Ni}  \;\stackrel{t_{1/2} = \; 6.08d}{\hbox to 60pt{\rightarrowfill}} \; ^{56}\mathrm{Co} \; 
\stackrel{t_{1/2} = \; 77.2d}{\hbox to 60pt{\rightarrowfill}} \; ^{56}\mathrm{Fe} \\*
& &^{57}\mathrm{Ni}  \;\stackrel{t_{1/2} = \; 35.60 h}{\hbox to 60pt{\rightarrowfill}}\; ^{57}\mathrm{Co} \;
\stackrel{t_{1/2} = \; 271.79d}{\hbox to 60pt{\rightarrowfill}} \; ^{57}\mathrm{Fe}\\*
& &^{55}\mathrm{Co}  \;\stackrel{t_{1/2} = \; 17.53 h}{\hbox to 60pt{\rightarrowfill}}\; ^{55}\mathrm{Fe} \;
\stackrel{t_{1/2} = \; 999.67 d}{\hbox to 60pt{\rightarrowfill}} \; ^{55}\mathrm{Mn.}
\end{eqnarray}
Because of the abundance variations between the two scenarios, the predicted bolometric luminosity $>1600$ days after $t_{B {\rm max}}$ differs by $>50\%$ between the SD and DD models \citep{roepke12}.  Thus, very-late-time bolometric observations of SNe Ia should be a strong diagnostic of these nuclear yields, and discriminate between progenitor models. 

However, observing SNe Ia at very-late times has proven extremely difficult. The latest observations of a SNe Ia before this work were long-pass ($F350LP$) observations of SN~2012cg taken by the {\it Hubble Space telescope} (HST) 1055 days after $t_{B {\rm max}}$. Through these observations, \citet{graur16} may have been the first to observe the decay of $^{57}\mathrm{Co}$ powering a SN Ia light curve, but their observations were also consistent with a light echo.
SN~2012cg was not observable at late enough times to constrain the presence of $^{55}\mathrm{Fe}$.

The recent SN Ia 2011fe is our best opportunity to observe a SN Ia at extremely late times.  SN~2011fe was discovered less than one day after the explosion by the Palomar Transient Factory \citep{law09} in a sparse region of the outer disk of the well-studied, face-on spiral M101. SN~2011fe became the brightest SN Ia in almost 40 years and exploded a mere 6.4 Mpc away \citep{shappee11}. Its early discovery allowed extensive multi-wavelength follow-up observations to be quickly carried out and these studies have shown that SN~2011fe was a ``plain vanilla'' SN Ia \citep{wheeler12}, remarkable only in its proximity.  \citet{li11} and \citet{graur14} used pre-explosion HST images to constrain the progenitor system of SN~2011fe. \citet{li11} did not detect any counterpart at the location of the SN and placed $2 \sigma$ upper limits of F435W $<27.87$, F555W $< 27.49$, and F814W $< 26.81$ at its position.  \citet{patat13} obtained high-resolution spectra of SN~2011fe and found that it is only slightly reddened and appears to be surrounded by a ``clean'' environment with a total reddening from both the Milky Way and M101 of $E(B-V) \lesssim 0.05$ mag.  Dust is expected to form in the ejecta of SNe~Ia $\sim 100$ -- $300$ days after $t_{B {\rm max}}$ \citep{nozawa11}.  However, there is no evidence for dust formation during the first 3 years \citep{kerzendorf14} and the optical depth would drop as $t^{-2}$, so dust extinction is unimportant for late-time observations of SN~2011fe.  Additionally, very-late-time ($\sim 1000$ days after $t_{B {\rm max}}$) spectroscopic observations showed no evidence of any light echoes (\citealp{graham15b}, \citealp{taubenberger15}; Shappee et al. in preparation).  These factors make SN~2011fe an exceptional opportunity to put the strongest constraints on the progenitor system of any SN Ia. While SN~2014J \citep{fossey14CBET} is still closer, it is not a good candidate for late-time studies because it is highly obscured (e.g., \citealp{brown15}), located in a crowded, dense environment, and is already showing evidence of light echoes \citep{crotts15}.

The data available for SN~2011fe are unparalleled, with multi-wavelength studies in the radio \citep{chomiuk12, horesh12}, the far-IR with Herschel \citep{johansson13}, the mid-IR with {\it Spitzer} \citep{mcclelland13}, the IR \citep{matheson12}, the optical \citep{nugent11, smith11, richmond12, munari12, shappee13, patat13, pereira13, tsvetkov13, kerzendorf14, graham15a, graham15b, lundqvist15, shappee15ATEL, zhang16, friesen17}, the UV with {\it Swift} \citep{nugent11, brown12} and HST \citep{foley13, mazzali14, friesen17}, X-rays with {\it Swift} \citep{horesh12, margutti12} and Chandra \citep{margutti12}, and gamma rays \citep{isern13}. By looking for observational signatures from the companion, many of these studies have already placed constraints on SNe Ia progenitor models. For example, \citet{bloom12}  do not detect the early shock interaction expected from the collision of the eject with a nearby companion \citep{kasen10}.  Additionally,  \citet{shappee13} do not detect the material expected to be stripped from a non-degenerate companion \citep{wheeler75}.  These studies have claimed to rule out the canonical SD channel where the companion is a normal red giant or main-sequence star. However, \citet{justham11} and \citet{distefano12} have suggested a way to avoid these limits using rotational support of the WD to delay the explosion long enough for the companion to evolve into a detached, smaller, and fainter star.  Alternatively, \citet{wheeler12} suggested that the companion could be a less massive M~dwarf with a smaller radius than previously considered.  These variations on the SD model highlight the need to have a direct probe into the properties of the exploding WD(s) instead of just relying on possible interactions with a companion for observational signatures.

In Section~\ref{sec:Obs}, we describe our observations.  In Section~\ref{sec:Var}, we constrain the pre-explosion variability at the site of SN~2011fe. In Section~\ref{sec:Bol}, we construct a late-time quasi-bolometric light curve of SN~2011fe and use it to place constraints on the progenitor system. In Section~\ref{sec:Contaminates}, we discuss possible sources of contamination in our quasi-bolometric light curve. Finally, we discuss our results in Section~\ref{sec:conclusion}. Throughout this paper we assume SN~2011fe is at a distance of $6.4\pm0.5$ Mpc \citep{shappee11}, has a Milky Way reddening of $E(B-V) = 0.009$ mag \citep{schlegel98} with  a ratio of total-to-selective absorption of $R_{V} = 3.1$ \citep{cardelli89}, and had no host-galaxy extinction (e.g., \citealp{patat13}).

\section{Observations}
\label{sec:Obs}

\begin{figure*}[htp]
	\centerline{\includegraphics[width=18.0cm]{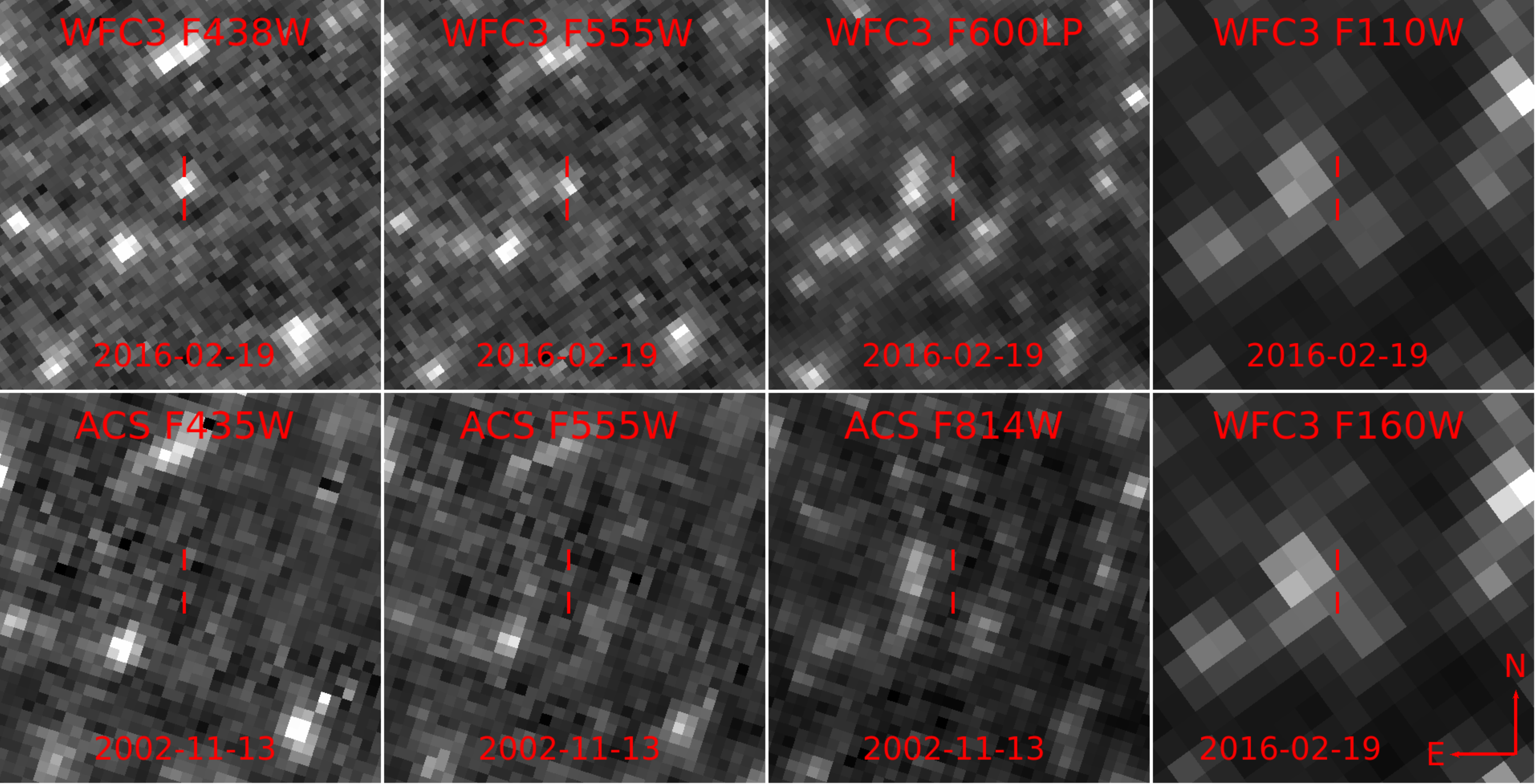}}
	\caption{Late-time ($\pastmax$ days after $t_{B {\rm max}}$) and archival pre-explosion (GO9490; PI: K. Kuntz) HST images of the location of SN~2011fe.  The tick marks are $0.\!\!''1$ long and indicate our measured position for SN 2011fe ($\textrm{RA}=14^{\rm h}03^{\rm m}05.\!\!^{\rm{s}}740$  $\textrm{Dec}=+54\degree16'25.\!\!''24$ J2000; astrometric coordinates referenced to the WCS solution of icoy01080\_drz.fits).}
	\label{fig:image}
\end{figure*}

\begin{figure*}[htp]
	\centerline{\includegraphics[width=18.0cm]{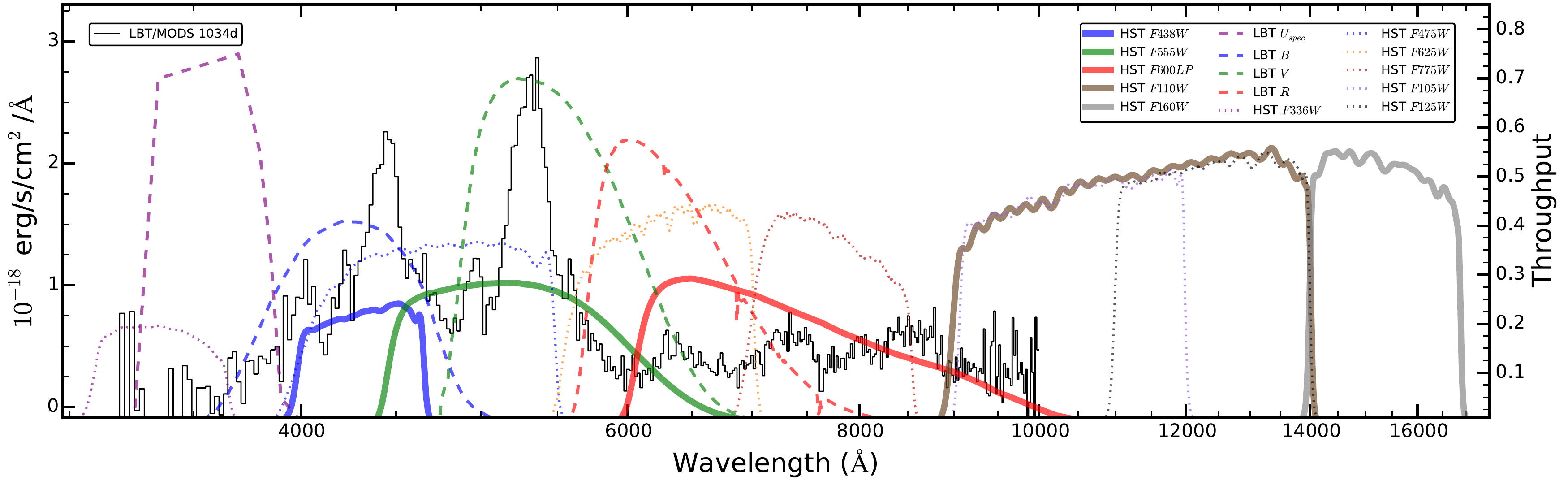}}
	\caption{Throughput functions for the filters used in this work as compared to the \citet{taubenberger15} MODS/LBT spectrum of SN~2011fe taken $1016$ days after $t_{B {\rm max}}$.}
	\label{fig:filters}
\end{figure*}

\begin{figure*}[htp]
	\centering
	\includegraphics[width=18.0cm]{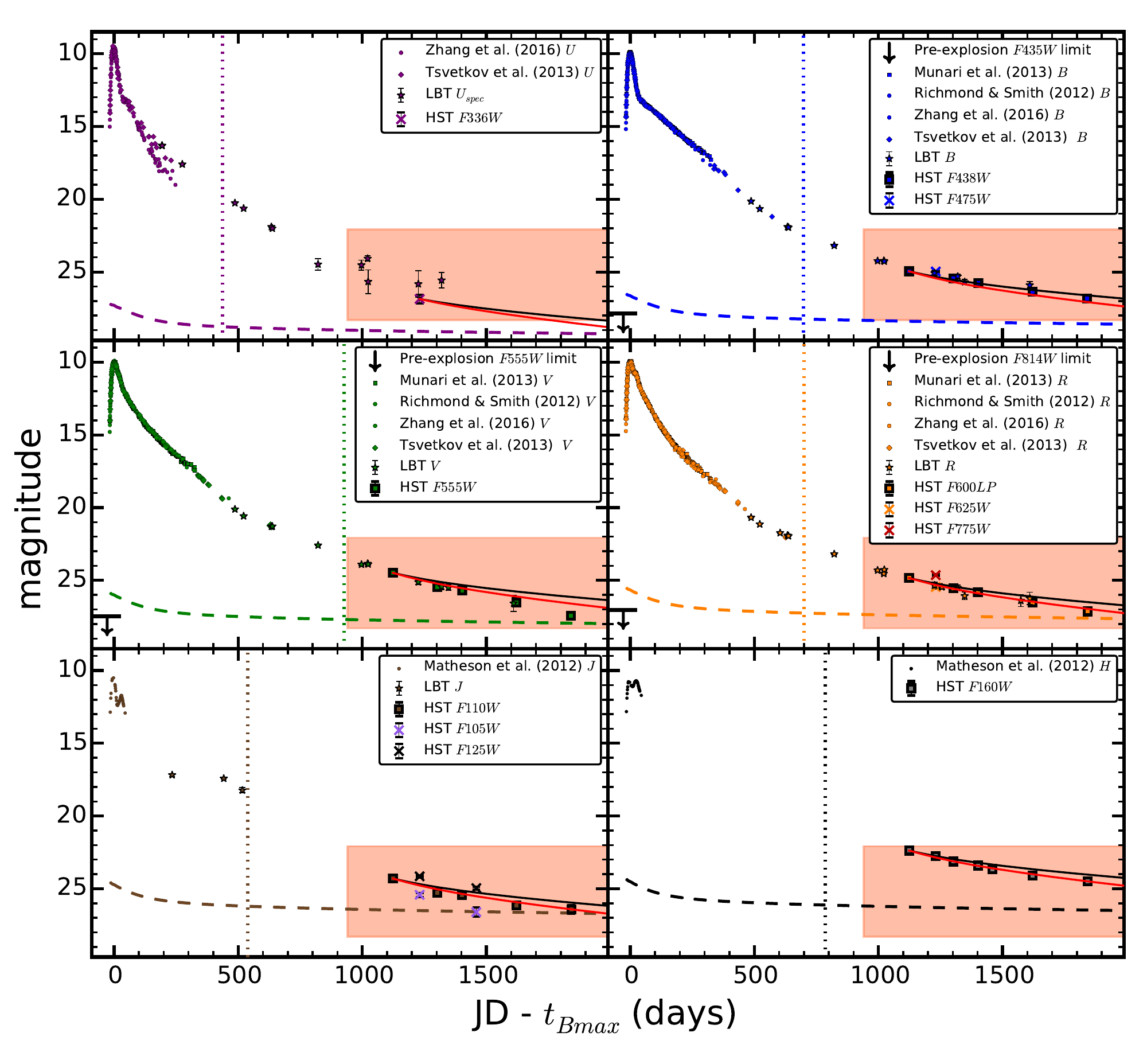}

	\caption{HST (filled squares and crosses), LBT (filled stars), and published (indicated in legends) light curves of SN~2011fe. The uncertainties are shown for the LBT and HST observations but can be smaller than the points. The vertical dotted lines show the previously latest SN Ia observations in each filter excluding examples of light echoes \citep{stritzinger07, leloudas09, cappellaro97}. The horizontal dashed lines show the expected brightness of a shock-heated 1 \msun{} MS companion \citep{shappee13b} for each filter.  The solid black and solid red lines show the \citet{roepke12} SD  and DD models, respectively, scaled to fit the first HST observation.  Black arrows just below $t=0$ days indicate the \citet{li11} 2-sigma pre-explosion upper limits. The orange highlighted region is shown in Figure~\ref{fig:lightcurve}.}
	\label{fig:lightcurvefull}
\end{figure*}

\begin{figure*}[htp]
	\centering
	\includegraphics[width=18.0cm]{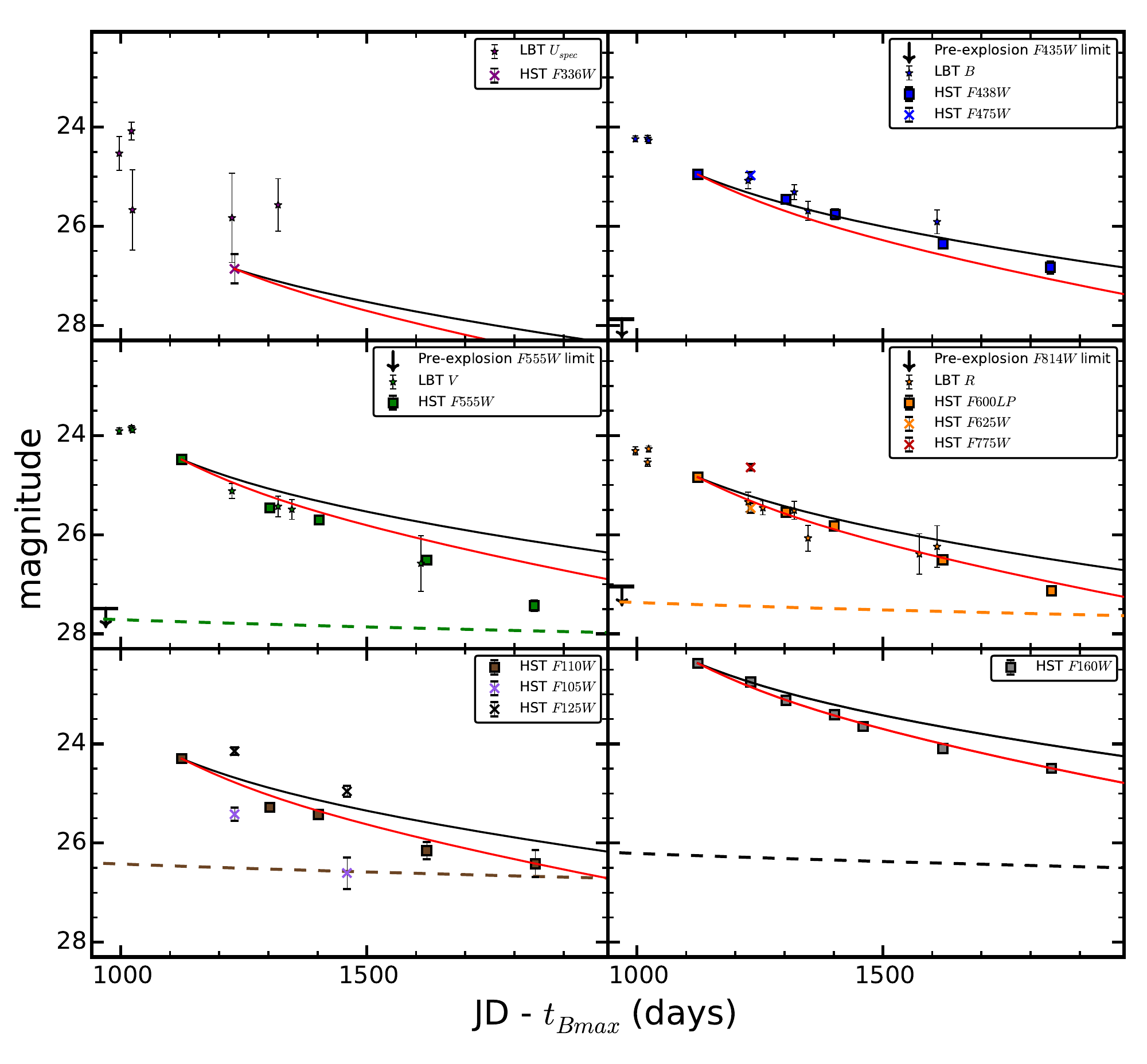}

	\caption{HST (filled squares and crosses) and LBT (filled stars) late-time light curves of SN~2011fe. The uncertainties are shown but are smaller than the points in some cases.  The horizontal dashed lines show the expected brightness of a shock-heated 1 \msun{} MS companion \citep{shappee13b} for each filter.  The solid black and solid red lines show the \citet{roepke12} SD  and DD models, respectively, scaled to fit the first HST observation.  The black arrows just below $t=1000$ days indicate the \citet{li11} 2$\sigma$ pre-explosion upper limits. }
	\label{fig:lightcurve}
\end{figure*}

We analyzed pre- and post-explosion images of SN~2011fe from the Large Binocular Camera (LBC; \citealp{giallongo08}) and the NIR spectrograph LUCI \citep{seifert03} on the Large Binocular Telescope (LBT) and the Wide Field Camera 3 (WFC3) and the Advanced Camera for Surveys (ACS) on HST. Our last epoch of WFC3 imaging and archival, pre-explosion ACS imaging at the location of SN~2011fe are shown in Figure~\ref{fig:image}. The SN is easily seen even in our last epoch.  In Figure~\ref{fig:filters} we show the throughput of the optical filters used in this work along with a spectrum of SN~2011fe acquired $1016$ days after $t_{B {\rm max}}$ using MODS on the LBT \citep{taubenberger15}. The photometry is presented in Table~\ref{tab:phot} and shown in Figures~\ref{fig:lightcurvefull} and \ref{fig:lightcurve}.

\begin{deluxetable}{lrcc}
\tablewidth{240pt}
\tabletypesize{\footnotesize}
\tablecaption{Photometric Observations}
\tablehead{
\colhead{JD} &
\colhead{Band} &
\colhead{Mag. | Bol. Flux} &
\colhead{Telescope} \\ 
\colhead{($-$2,450,000)} &
\colhead{} &
\colhead{\ \ \ \ \ \     \nodata | [$10^{36}$ ergs/s]}  &
\colhead{}  }
\startdata
6008.836 & $U_{\textrm{spec}}$ & 16.31(0.01) & LBT/LBC \\ 
6302.987 & $B$ & 20.14(0.02) & LBT/LBC \\ 
6303.001 & $V$ & 20.11(0.02) & LBT/LBC \\ 
6302.993 & $R$ & 20.68(0.02) & LBT/LBC \\ 
6050.000 & $J$ & 17.18(0.10) & LBT/LUCI \\ 
6939.664 & $F438W$ & 24.96(0.08) & HST/WFC3/UVIS \\ 
6939.653 & $F555W$ & 24.48(0.03) & HST/WFC3/UVIS \\ 
6939.667 & $F600LP$ & 24.84(0.03) & HST/WFC3/UVIS \\ 
6939.664 & $F110W$ & 24.30(0.05) & HST/WFC3/IR \\ 
6939.789 & $F160W$ & 22.37(0.02) & HST/WFC3/IR \\ 
6939.664 & bol & 13.74(0.22) & HST/WFC3 \\ 
\enddata \tablecomments{Photometric and bolometric light curves. Photometry is calibrated in the Vega magnitude system  and the bolometric light curve is in units of $10^{36}$ ergs/s. \textit{Only the first observation in each band is shown here to demonstrate its form and content. The table is included in its entirety as an ancillary file.}} 
\label{tab:phot} 
\end{deluxetable}

We obtained 28 epochs of LBT/LBC $U_{\textrm{spec}}-$, $B-$, $V-$, and $R-$band imaging of M101 between 2008 March and 2016 February as part of a program searching for failed SNe \citep{kochanek08, gerke11, gerke15}. The pre-explosion images allowed us to construct deep reference images without having to wait for SN~2011fe to completely fade. The median full width at half maximum in the  $U_{\textrm{spec}}-$, $B-$, $V-$, and $R-$band images containing SN~2011fe were 1.\!\!''27, 1.\!\!''22, 1.\!\!''11, and 1.\!\!''06, respectively. The data were analyzed with the {\tt ISIS} image-subtraction package \citep{alard98, alard00}.  We then performed aperture photometry on the subtracted images using the IRAF {\tt apphot} package and calibrated the magnitudes using the SDSS Data Release 10 (DR10; \citealp{ahn14}) photometry of nearby fields stars transformed into Bessel filters using Lupton (2005)\footnote{\url{http://www.sdss.org/dr12/algorithms/sdssUBVRITransform/\#Lupton2005}}.  We used 0.\!\!''7, 0.\!\!''9, 0.\!\!''9, and 0.\!\!''9 radius apertures for the $U_{\textrm{spec}}-$, $B-$, $V-$, and $R-$band photometry, respectively.  We note that the \citet{zhang16} $U$-band observations disagree with our first two $U_{\textrm{spec}}$ epochs and the \citet{tsvetkov13} observations.



We also obtained 3 epochs of LBT/LUCI $J-$band imaging of SN~2011fe between 2012 May and 2013 February.  The photometry was obtained from $J-$band acquisition images for LUCI spectroscopy \citep{mazzali15}. Dithered pairs of J-band images were subtracted and aperture photometry was applied to the supernova and to three 2MASS catalog stars in the field for calibration.

We also obtained four epochs of HST data using the WFC3 UVIS and IR cameras (GO-13737 and 14166) starting in 2014 October.  We used the WFC3/UVIS $F438W$ ($B$), $F555W$ ($V$), $F600LP$ ($R + I$), $F110W$ ($Y + J$), and $F160W$ ($H$) filters because the very-late-time spectra of SNe Ia predominantly consist of line emission over a very broad range of wavelengths redward of the UV (see Figure~\ref{fig:filters}).    This filter combination was selected for several reasons. (1) The filters cover 4000 to 17000\AA\ with no gaps and relatively uniform throughput. Ground-based spectroscopy and photometry showed there is little flux blueward of 4000 \AA{} so we did not include the F336W ($U$) or UV filters.  (2) A single (or two) long-pass filter(s) would not be sufficient because the shape of the SN spectral energy distribution (SED) is known to be changing with the ionization state (\citealp{mcclelland13, graham15b, taubenberger15}, Shappee et al. in preparation).  Computing the bolometric luminosity with a single F350LP observation would lead to large systematic uncertainties because we could mistake a change in ionization state for a change in the bolometric luminosity or interpret a light echo as a flattening of the bolometric light curve.   (3) Finally, this optical filter combination evenly partitions the emission seen in the $1016$-day spectrum obtained by \citet{taubenberger15} across the filters.  

We measured the instrumental Vega Magnitudes using the WFC3 module of the {\tt DOLPHOT} stellar photometry package\footnote{\url{http://americano.dolphinsim.com/dolphot/}} for our WFC3 observations. For each of the UVIS images, we first corrected for charge-transfer efficiency (CTE) at the pixel level using the WFC3/UVIS CTE tool \footnote{\url{http://www.stsci.edu/hst/wfc3/tools/cte\_tools}}. The NIR was not as simple as the optical because SN~2011fe is blended with nearby, very-red stars in the NIR (see Figure~\ref{fig:image}). First, we determined the precise position of SN~2011fe and the nearby stars in the $F110W$ and $F160W$ images by using the higher-resolution $F600LP$ images.  With the recommended parameters, {\tt DOLPHOT} correctly found the nearby stars and SN~2011fe for the $1124$ and $1403$ day epochs. However, for the $1302$,  $1622$, and $1840$ day epochs we had to adjust the source thresholds and maximum iterations for {\tt DOLPHOT} to correctly identify these sources.  Finally, we verified the flux differences between epochs by subtracting epochs with {\tt ISIS}.  SN~2011fe is clearly visible in these subtractions and the flux differences are consistent with our reported {\tt DOLPHOT} photometry.

However, \citet{williams14} found that the photon statistics errors reported by {\tt DOLPHOT} for the Panchromatic Hubble Andromeda Treasury survey tended to underestimate uncertainties by a factor of a few.  We corrected the uncertainties of our HST photometry separately for the optical and NIR by analyzing the {\tt DOLPHOT} light curves for point sources within $18.\!\!''0$ of the SN.  We required that sources were detected in all four epochs within the magnitude range observed for SN 2011fe in our HST data.  We then removed high-amplitude variable stars, defined by those with Stetson $F555W/F600LP$ or $F110W/F160W$ variability indices $I > 10$ \citep{stetson96}, leaving $149$ stars in the optical and $262$ stars in the NIR to calibrate our errors.  For each star we determined the ratio of the deviation from the star's mean magnitude to the error reported by {\tt DOLPHOT} in each epoch.  We then scaled the {\tt DOLPHOT} errors of SN~2011fe by the median of these ratios for each filter and epoch separately.  All uncertainties were scaled by less than a factor of $2.5$, except for the $F438W$ filter's $1403$ day epoch which was scaled by a factor of $3.2$.  

Finally, for completeness, we also present archival HST WFC3/UVIS, ACS, and WFC3/IR photometry in the F336W ($U$), F475W ($g$), F625W ($r$), F775W ($i$), F105W ($Y$), F125W ($J$) and F160W ($H$) filters acquired during 2015 January (GO-13824; PI W. Kerzendorf).  We treat the WFC3 data as described above and we perform photometry on the ACS data using the ACS module of the {\tt DOLPHOT} stellar photometry package.  We did not scale this photometry's errors because only a single epoch of observation is available.   This photometry is consistent with ours, as seen in Figure~\ref{fig:lightcurve}.

\section{Pre-explosion LBT Variability Constraints}
\label{sec:Var}

We have 7, 9, 9, and 8 epochs of LBT/LBC data obtained between 2008 March and 2010 December observed in $U$, $B$, $V$, and $R$, respectively.  \citet{gerke14ATEL} previously used a similar dataset of M82 to set a limit on the pre-explosion $R$-band variability at the location for SN~2014J of $\Delta | \nu L_{\nu} | < 3000$ \Lsun{} ($3 \sigma$) over the 6 years prior to the SN.  We observe no pre-explosion variability at the location of SN~2011fe and we place 3-sigma limits on the pre-SN variability in $U$, $B$, $V$, and $R$ of \varUlfl{}, \varBlfl{}, \varVlfl{}, and \varRlfl{} erg s$^{-1}$ or $\sim \varUbolLsun{}$, $\sim \varBbolLsun{}$, $\sim \varVbolLsun{}$, and $\sim \varRbolLsun{}$ \Lsun{}, respectively.  Interpolating linearly between the filters and extrapolating using a constant $f_{\lambda}$, we find a variability limit of \varbol{} erg s$^{-1}$ or $\sim \varbolLsun{}$ \Lsun{} over the wavelength range from $\varwavedw$ -- $\varwaveup$ \AA{} .  
These variability limits are approximately an order of magnitude weaker than the pre-explosion upper limits placed on the progenitor system from archival HST observations \citep{li11} and only rule out relatively large outbursts from the progenitor system.

\section{Quasi-Bolometric Light Curve and Progenitor Constraints}
\label{sec:Bol}

In this section we construct a quasi-bolometric light curve for SN~2011fe, place constraints on the progenitor assuming our quasi-bolometric light curve is representative of the true bolometric light curve, and compute the fraction of the bolometric luminosity emitted in our observed wavelengths.

\subsection{Quasi-Bolometric Light Curve}
\label{sec:BolLC}

\begin{figure*}[htp]
	\centerline{\includegraphics[width=18.0cm]{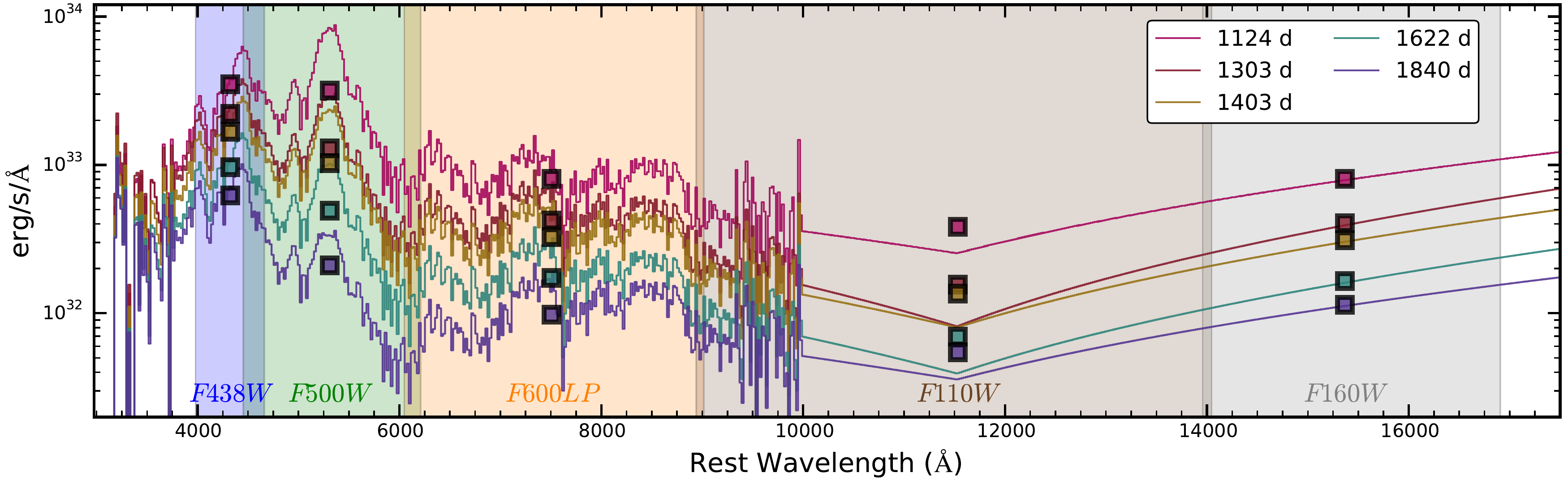}}
	\caption{Late-time SED for SN~2011fe.  The squares show our HST observations at their effective wavelengths.  The solid lines show the fit SED as described in Section~\ref{sec:BolLC}. Synthetic photometry of the SED are not expected to go through the HST photometry, but instead agree to better than \stopit{} mag with the filter-averaged values in each filter.  The colors of the symbols and lines distinguish between epochs. The shaded regions indicate where each filter's throughput is $> 10 \%$.}
	\label{fig:SED}
\end{figure*}

\begin{figure*}[htp]
	\centering
	\includegraphics[width=15.0cm]{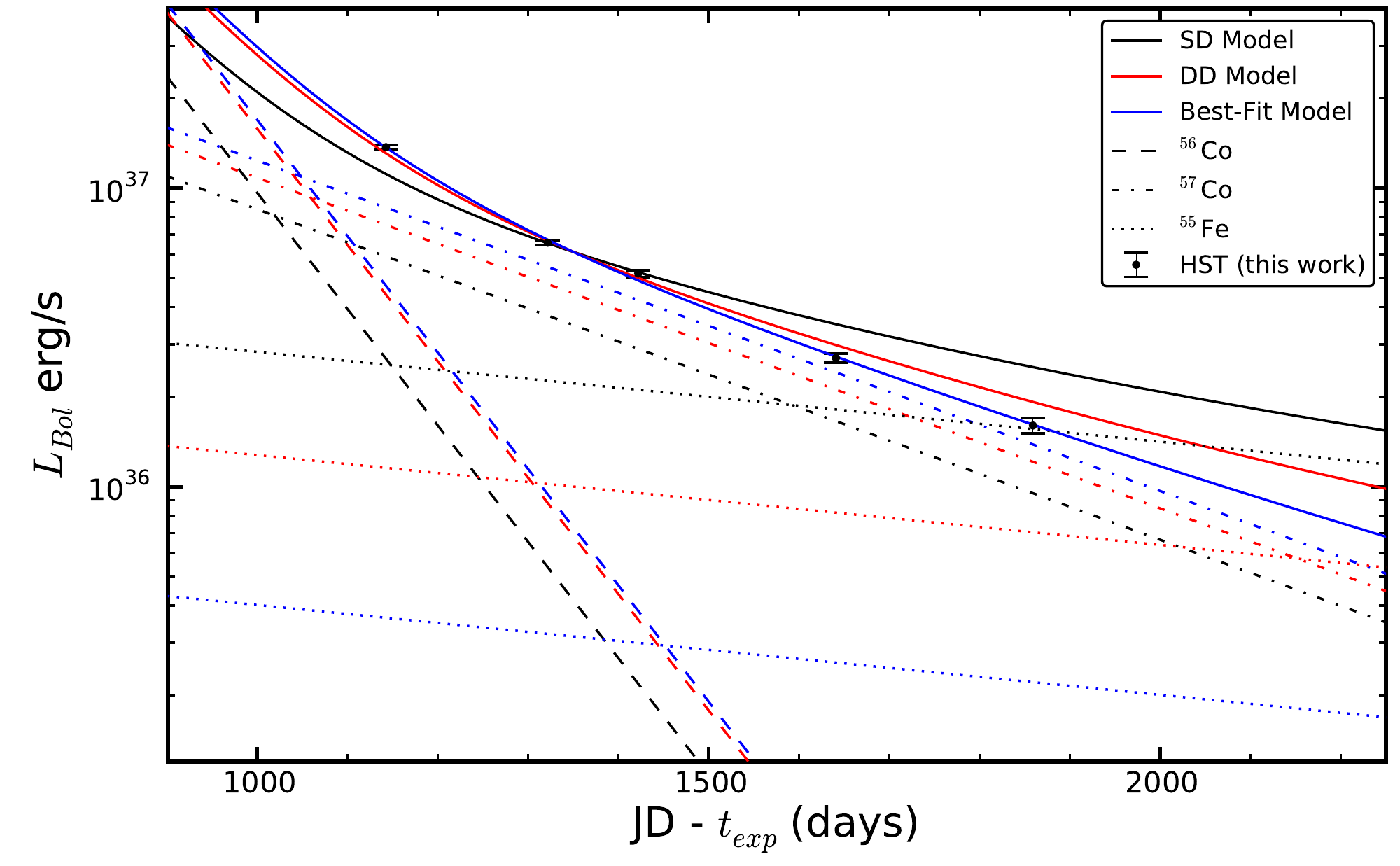}
	\includegraphics[width=15.0cm]{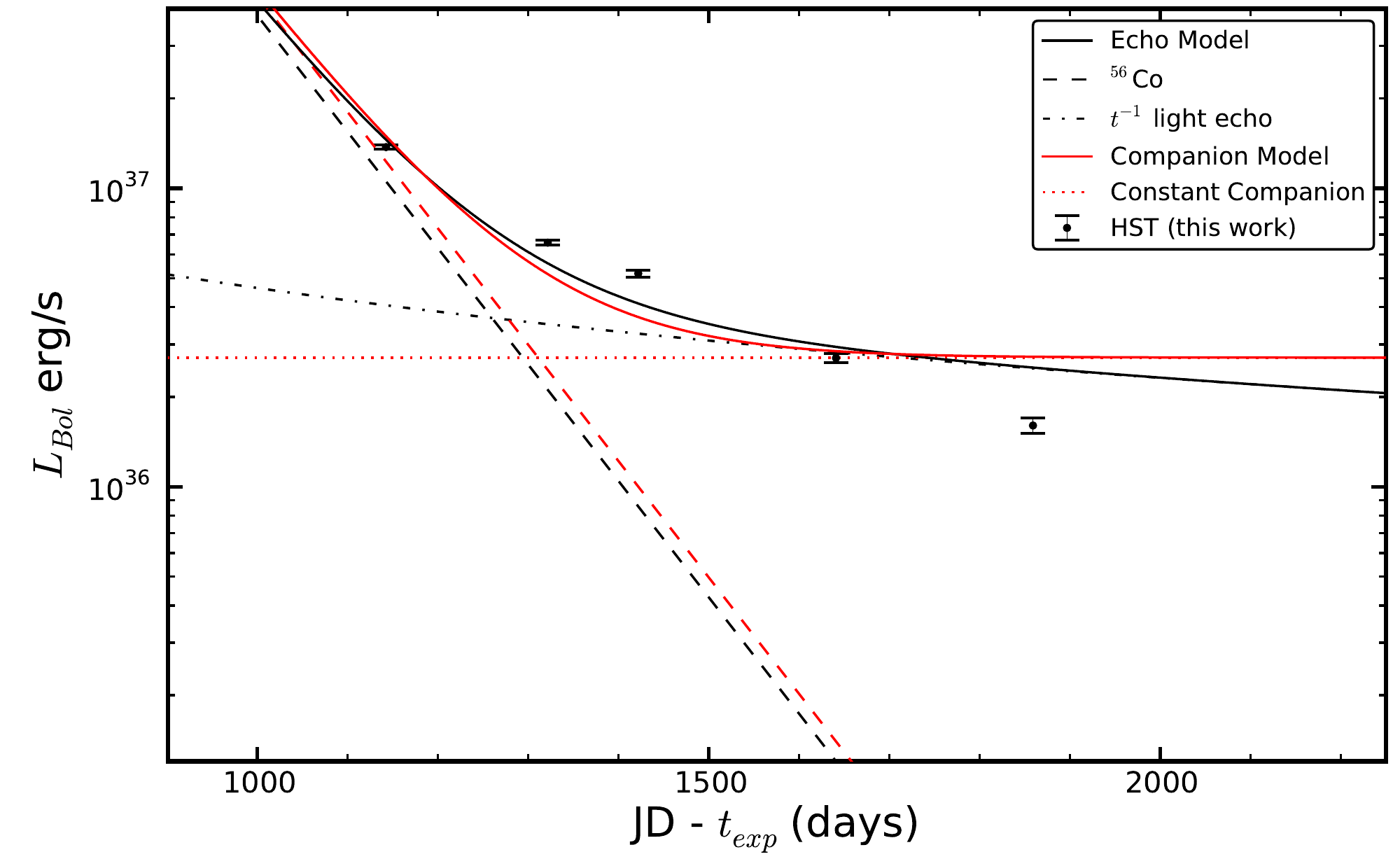}

	\caption{The optical+NIR quasi-bolometric light curve of SN~2011fe from $\bolwavedw$ -- $\bolwaveup$ \AA{}. {\bf Top Panel:} the red and blue lines are the scaled SD and DD models from \citet{roepke12}, respectively. The blue lines are the best-fit model treating $^{56}\mathrm{Ni}$, $^{57}\mathrm{Co}/^{56}\mathrm{Ni}$ and $^{55}\mathrm{Fe}/^{57}\mathrm{Co}$ as free parameters. {\bf Bottom Panel:} the black lines are the best-fit light echo model including $^{56}\mathrm{Ni}$ decay and a $t^{-1}$ light echo. The red lines are the best-fit companion model including $^{56}\mathrm{Ni}$ decay and a constant flux from a surviving SD companion.}
	\label{fig:bol}
\end{figure*}

The quasi-bolometric light curve for SN~2011fe was computed over the wavelength region that our photometric observations probe, from $\bolwavedw$ -- $\bolwaveup$ \AA{}.
To reduce the systematic errors in the calculation, we did not assume the optical SED was a simple interpolation over the photometric bands, but instead used the \citet{taubenberger15} $1016$-day MODS/LBT spectrum of SN~2011fe as a template and iteratively fit it to the observed photometry. For the NIR there exist no published late-time spectra.  Instead we averaged $f_{\lambda}$ from $9000$ -- $10000$ \AA{} in the $1016$-day MODS/LBT spectrum and assumed a flat $f_{\lambda}$ spectrum from $10000$ -- $\bolwaveup$ \AA{}.  We note that even in the worst case, where the underlying SED shifts from one edge of the NIR filters to the other, only a $\sim 10\%$ error is introduced in our bolometric light curve. First, we performed synthetic photometry from this constructed spectrum for the $F438W$, $F555W$, $F600LP$, $F110W$, and $F160W$ bands and computed the differences between the synthetic and observed photometry.  Second, we linearly interpolate these differences in between  each filter's effective wavelength.  Finally, we multiplied the spectrum by these differences.  We iterated this process until the synthetic and observed photometry matched to better than \stopit{} mag in each filter. The warped SEDs for each epoch are shown in Figure~\ref{fig:SED}.  The errors on the quasi-bolometric light curve were computed through Monte Carlo resampling the input photometry and then recomputing the quasi-bolometric light curve.  The quasi-bolometric light curve for SN~2011fe is shown in Figure \ref{fig:bol} and presented in Table~\ref{tab:phot}.

Finally, because SN~2011fe is crowded in the NIR, its photometry might be questionable and affect our analysis based on the quasi-bolometric light curve.  To alleviate this concern, we also repeated the quasi-bolometric light curve calculation using only the optical data ($\bolwavedwopt$ -- $\bolwaveupopt$ \AA{}) where SN~2011fe is well isolated and pre-explosion observations show no sources to limits fainter than SN~2011fe in our last epoch. We then duplicated the entire analysis that follows and obtained qualitatively similar results.

\subsection{Progenitor Constraints}
\label{sec:Progenitor}

We constrain the progenitor models by assuming the quasi-bolometric light curve computed in Section~\ref{sec:Bol} tracks the true bolometric decay of SN~2011fe (see Figure \ref{fig:bol}).  
\citet{milne01} showed that this is likely a good assumption at earlier times, between $\sim 50$ and $600$ days, where the $V$-band scales with the bolometric luminosity.    
We fit the quasi-bolometric light curve using Bateman's equation for the radioactive-decay chains of $^{56}\mathrm{Ni}$, $^{57}\mathrm{Ni}$ and $^{55}\mathrm{Co}$ presented in Equations~1-3.  We assume the decay energies presented in Table 1 of \citet{seitenzahl09}; that the SN ejecta is transparent to $\gamma$-rays; that $e^{-}$, $e^{+}$, and X-rays instantaneously deposit their energy; and that deposited energy is immediately radiated.  Section~3 of \citet{graur16} provides a discussion and justification of these assumptions. 

First, we fit the \citet{roepke12} predicted abundances for $^{56}\mathrm{Ni}$, $^{57}\mathrm{Ni}$ and $^{55}\mathrm{Co}$ from their SD and DD models to the bolometric light curve. Because we cannot observe SN~2011fe at all wavelengths we are not able to make true calorimetric measurements of SN~2011fe and thus we cannot make absolute abundance measurements from our quasi-bolometric light curve.  However, we are able to make relative abundance measurements by assuming the quasi-bolometric light curve is proportional to the true bolometric light curve.  We found that the quasi-bolometric light curve of SN~2011fe strongly prefers the DD model ($\chisq = \VMChiSq{}$ with $\nu = \VMdof$ degrees of freedom) over the SD model ($\chisq = \DDChiSq$ with $\nu = \DDdof$).  The majority of this $\chisq{}$ difference is due to the data preferring the lower $^{55}\mathrm{Fe}/^{57}\mathrm{Co}$ ratio predicted from nuclear burning at lower densities. 

We used the {\tt emcee} package \citep{foreman13}, a Python-based implementation of the affine-invariant ensemble sampler for Markov chain Monte Carlo (MCMC), to fit the quasi-bolometric light curve with the ratios $^{57}\mathrm{Co}/^{56}\mathrm{Ni}$ and $^{55}\mathrm{Fe}/^{57}\mathrm{Co}$ as free parameters.  We found $\log (^{57}\mathrm{Co}/^{56}\mathrm{Ni})$ and $\log (^{55}\mathrm{Fe}/^{57}\mathrm{Co})$ abundance ratios of \CotoNifit{} and \FetoCofit{}, respectively.  Our best fit  had a $\chisq = \fitallChiSq{}$ with $\nu = \fitalldof$ for abundance ratios of  $^{57}\mathrm{Co}/^{56}\mathrm{Ni} = \CotoNibestfitval{}$ and $^{55}\mathrm{Fe}/^{57}\mathrm{Co} = \FetoCobestfitval{}$.  Finally, we fit the quasi-bolometric light curve excluding either $^{57}\mathrm{Co}$ or $^{55}\mathrm{Fe}$, finding $\chisq = \noCoChiSq{}$ and $\chisq = \noFeChiSq{}$, respectively, for $\nu = \noFedof$.  In practice, the present light curve can be explained using only $^{56}\mathrm{Ni}$ and $^{57}\mathrm{Co}$ with an upper limit on the $^{55}\mathrm{Fe}$ abundance ratio of $^{55}\mathrm{Fe}/^{57}\mathrm{Co} < \FetoCofitupper$ at 99\% confidence.  This upper limit strongly favors the \citet{roepke12} DD model ($^{55}\mathrm{Fe}/^{57}\mathrm{Co} = 0.27$) over their SD model ($^{55}\mathrm{Fe}/^{57}\mathrm{Co} = 0.68$). 

Using only the optical data to construct the bolometric light curve leads to similar results.  For the model with all the isotope masses allowed to vary we find $\log (^{57}\mathrm{Co}/^{56}\mathrm{Ni}) = \CotoNifitopt{}$ and $\log (^{55}\mathrm{Fe}/^{57}\mathrm{Co}) = \FetoCofitopt{}$.  We also find that the models require the presence of $^{57}\mathrm{Co}$ but not $^{55}\mathrm{Fe}$, constraining $^{55}\mathrm{Fe}/^{57}\mathrm{Co} < \FetoCofitupperopt$ at 99\% confidence.

\subsection{Fraction of the Total Bolometric Luminosity emitted in the Optical/NIR}
\label{sec:Fraction}

Although we cannot determine the absolute abundance of $^{56}\mathrm{Ni}$, we can use previous estimates of $M_{^{56}\mathrm{Ni}}$ for SN~2011fe to determine the fraction of the decay energy, excluding $\gamma$-rays, which is encompassed by our quasi-bolometric light curve. There have been two previous studies that have measured the total $^{56}\mathrm{Ni}$ yield for SN~2011fe.  \citet{pereira13} used an early-time UltraViolet Optical near-InfraRed (UVOIR) bolometric light curve and Arnett's rule (e.g., \citealp{stritzinger06}) and found $M_{^{56}\mathrm{Ni}} = 0.53 \pm 0.11$ \msun{}.  
Finally, \citet{mazzali15} used optical and near-infrared nebular spectra to estimate that $M_{^{56}\mathrm{Ni}} = 0.47 \pm 0.05$ \msun{}.  The weighted mean of these measurements, assuming their random errors are independent, is $M_{^{56}\mathrm{Ni}} = \MNiassume{}$ \msun{}.  

In our fit to the quasi-bolometric light curve with the abundances of $^{56}\mathrm{Ni}$, $^{57}\mathrm{Ni}$, and $^{55}\mathrm{Co}$ as free parameters, we find that $M_{^{56}\mathrm{Ni}} = \MNifit$ \msun{} is needed to explain the observed flux. This implies that  \BolinOptNIR{} of the bolometric luminosity is being emitted from $\bolwavedw$ -- $\bolwaveup$ \AA{} independent of the primary uncertainty in the luminosity (the distance).  This estimate is dominated by our first epoch where $^{56}\mathrm{Ni}$ still dominates the emission.   Combining the early-time measurement of $M_{^{56}\mathrm{Ni}}$ and our measured $^{57}\mathrm{Co}/^{56}\mathrm{Ni}$ we find $M_{^{57}\mathrm{Co}} = \MCoEst$ \msun{}, in tension with the \citet{fransson15} estimate of $\sim 0.02$ \msun{}.   Furthermore,  finding that \BolinOptNIR{} of the emission is in the optical/NIR is substantially higher than the $\sim 20\%$ predicted by \citet{fransson15}.  Even if we just measure the optical emission ($\bolwavedwopt$ -- $\bolwaveupopt$ \AA{}), it represents \Bolinopt{} of the bolometric emission expected from $M_{^{56}\mathrm{Ni}} = \MNiassume{}$ \msun{}.

\section{Sources of Possible Contamination}
\label{sec:Contaminates}

In this section, we discuss a surviving binary companion and light echoes as possible sources of contamination to the bolometric light curve.  

\subsection{Contamination by a binary companion?}
\label{sec:Companion}

\citet{bloom12} did not see any signs of a shock interaction between the SN ejecta and a companion in the early-time light curve of SN~2011fe, which places strict limits on the radius of a potential companion ($\lesssim 0.1$ \rsun).  However, this limit can be avoided for unfavorable viewing angles \citep{kasen10},  M~dwarf companions \citep{wheeler12}, or rapidly rotating WD primaries \citep{justham11, distefano12}.  Using archival pre-explosion HST imaging, \citet{li11} placed significantly deeper flux limits for any source at the location of SN~2011fe than we observe in our last HST epoch (see Figure~\ref{fig:image}).  These observations ruled out the existence of a giant companion and $\gtrsim 6.0$ \msun{} main-sequence companions.   Finally, using deep-spectroscopic Keck observations of SN~2011fe 963 days after $t_{B {\rm max}}$, \citet{graham15b} show that any shock-heated companion must have $T \gtrsim 10^{4}$ K and $L \lesssim 10^{4}$ \Lsun{}. Thus, if there is a companion still present, it was initially smaller than a giant and then was shock heated by the SN ejecta.  However, a 1 \msun{} main-sequence shock-heated companion should still be significantly fainter than the last epoch presented here (\citealp{shappee13b}; See Figure~\ref{fig:lightcurve}). 

A shock-heated companion will evolve on the Kelvin-Helmholtz time of the envelope ($\sim 10^3$--$10^4$ years; \citealp{shappee13b}).  Thus, we can approximate a potential companion as a constant source of flux because our observations span a much shorter time-scale.  We find that a model including just $^{56}\mathrm{Ni}$ decay and fitting the flux from a possible shock-heated companion (red lines in the bottom panel of Figure~\ref{fig:bol}) cannot adequately describe the observed light curve ($\chisq = \compChiSq{}$ with $\nu = \compdof$).  If we fit the quasi-bolometric light curve including $^{56}\mathrm{Ni}$ and $^{57}\mathrm{Co}$ radioactive decay and a constant flux companion, we find that a constant flux source can only contribute $< \Compfitupper$ erg s$^{-1}$ ($\sim \CompfitupperLsun{}$ \Lsun{}) to our quasi-bolometric light curve at 90\% confidence.

\subsection{Light Echoes?}
\label{sec:Echo}

As previously mentioned, SN~2011fe is observed to have exploded into an extremely clean environment.  However, the light from SN~2011fe has now decreased by more than a factor of $\sim 7$ million in flux, so even faint light echoes could significantly contribute to the observed light curves.  

As a first step, we searched for large-scale light echoes from SN~2011fe that would not affect our photometry but would allow us to probe the interstellar medium in M101 in the vicinity of SN~2011fe. To estimate the largest angular scale a light echo from SN~2011fe could have during our fourth HST epoch we conservatively assume a scattering screen $1$ kpc from SN~2011fe and use the law of cosines to find a maximum angular scale of $1.\!\!''7$ in radius.  We used the $F438W$ images to search the region around SN~2011fe because dust tends to scatter blue light more efficiently than red.  We subtracted all the epochs from the image acquired 1403 days after $t_{B {\rm max}}$.   Ten to fifteen stars in common between images were used to transform the template image to the coordinates of the search epoch. The subtracted images were convolved with a 2-pixel Gaussian kernel to bring out low surface brightness residuals.  The subtractions reveal many variable point sources in the galaxy and there is a bright artifact in the template that is likely due to a bright star off the field of view. There are also trails along the columns that are most obvious in the shorter exposures and probably due to CTE. However, there are no obvious light echoes that have changed their position over the span of a year. Using the CTE trails from the first epoch to set a conservative upper limit, we see no light echoes within a very conservative $\sim 30''$ of SN~2011fe with a surface brightness greater than $2\times 10^{-3}$ $\mu$Jy~square-arcsec$^{-1}$.

We also searched for resolved light echos in close proximity of SN~2011fe.  Echoes from circumstellar material could be resolved in our last epoch with a maximum physical scale of about $ 2 \times c \times t \simeq 3.1$ pc and an angular size of $0.\!\!''10$ in diameter.   In the last epoch, the nearest two sources are both 4.8 pixels ($0.\!\!''19$) away from the position of SN~2011fe.  Both sources are red and are classified by {\tt DOLPHOT} as point sources with magnitudes of $F438W = 30.2 \pm 1.9$, $F555W = 27.7 \pm 0.2$, and $F600LP = 26.05 \pm 0.05$ and $F555W = 27.7 \pm 0.2$ and $F600LP = 25.72 \pm 0.04$.  The second source is not detected at $F438W$. Both sources are present in all of the earlier HST epochs, including those prior to the SN, so neither is a light echo.  SN~2011fe itself is classified as a point source by {\tt DOLPHOT} in every epoch.  As part of the photometric analysis, {\tt DOLPHOT} produces a residual map after subtracting a model of all the detected  sources. We visually inspect the residual maps around the location of SN~2011fe and we see no resolved structures. Thus, there is no evidence for a resolved light echo of SN~2011fe in our data that would affect our photometry of SN~2011fe.  

It is possible, however, to have an unresolved light echo along our line of sight.  Unresolved light echoes are expected to fade as $t^{-1}$ (see \citealp{graur16}).  If we model the quasi-bolometric light curve including only $^{56}$Co decay and an unresolved, $t^{-1}$ light echo we obtain a poor fit ($\chisq = \echoChiSq$ with $\nu = \echodof$; see Figure~\ref{fig:bol}).  Thus, our detection of $^{57}$Co decay cannot be explained as a light echo.  For a fit to the quasi-bolometric light curve including $^{56}\mathrm{Ni}$ and $^{57}\mathrm{Co}$ radioactive decay along with a $t^{-1}$ light echo, we find that the echo can only contribute $< \echofitupper$ erg s$^{-1}$ ($\sim \echofitupperLsun{}$ \Lsun{}; $\sim \echofitupperPer \%$ ) to our quasi-bolometric light curve at the last epoch ($1840$ days after $t_{B {\rm max}}$) at 90\% confidence.  Perhaps the strongest evidence against an unresolved light echo, however, is the color evolution of SN~2011fe.  At maximum light, SN~2011fe had $B-H = \BmHmax$.  $\pastmax$ days after $t_{B {\rm max}}$, SN~2011fe had $F438W - F160W = \BmH$.  Light echoes are expected to be roughly the same color or bluer than the SN at its peak \citep{patat05}, and SN~2011fe becoming redder by a factor of $\sim \BmHfact$ strictly limits the flux any light echo could be contributing.

%
%

\section{Discussion}
\label{sec:conclusion}

SN~2011fe has presented us with an exceptional opportunity to constrain SNe Ia progenitor models. A series of studies have already ruled out numerous models for the progenitor system of SN~2011fe based on interactions of the SN ejecta with a non-degenerate companion (e.g., \citealp{bloom12, shappee13}) or limits on any possible companion from pre-explosion imaging (e.g., \citealp{li11, graham15b}).  In this study, we add a novel and independent constraint.  At very-late times ($> 1050$ days after $t_{B {\rm max}}$), $^{57}\mathrm{Co}$ and then $^{55}\mathrm{Fe}$ become the dominant power sources for SN Ia light curves \citep{seitenzahl09} and the production of these isotopes is very sensitive to the central density of the exploding white dwarf(s) \citep{roepke12}. Thus, very-late-time bolometric observations of SNe Ia can distinguish between the SD and DD models. Using HST and LBT, we followed the light curve of SN~2011fe for an unprecedented $\pastmax$ days past $t_{B {\rm max}}$ and over a factor of $\sim 7$ million in flux.  We present the first clean detection of $^{57}$Co powering a SN Ia light curve and estimate a mass ratio of $\log (^{57}\mathrm{Co}/^{56}\mathrm{Co}) = \CotoNifit$.  The $^{57}\mathrm{Co}/^{56}\mathrm{Co}$ abundance ratio mildly prefers the \citet{roepke12} DD models ($\log (^{57}\mathrm{Co}/^{56}\mathrm{Co}) = -1.62$) over the SD models ($\log (^{57}\mathrm{Co}/^{56}\mathrm{Co}) = -1.51$).  We do not have a clean detection of a contribution from $^{55}\mathrm{Fe}$, finding a limit of $^{55}\mathrm{Fe}/^{57}\mathrm{Co} < \FetoCofitupper$ with 99\% confidence.  The $^{55}\mathrm{Fe}/^{57}\mathrm{Co} $ abundance ratio strongly prefer the lower central density of DD models ($^{55}\mathrm{Fe}/^{57}\mathrm{Co} = 0.27$) over the higher central density of SD models ($^{55}\mathrm{Fe}/^{57}\mathrm{Co} = 0.68$).

Theoretically, an infrared catastrophe (IRC) was predicted for SNe Ia beyond $\sim 500$ days when, as a result of cooling of the ejecta below $\sim 2000$ K, emission would transition from optical/NIR [Fe II-III] lines to the mid- and far-infrared fine-structure transitions \citep{axelrod80}.  If an IRC occurs, it would not be possible to observe very-late-time SNe Ia light curves.  Luckily, late-time observations of SNe Ia have not seen this transition (e.g., \citealp{sollerman04, leloudas09, mccully14b, kerzendorf14, graur16}). \citet{fransson15} claimed to reconcile an IRC with late-time observations of SNe Ia through the redistribution of UV emission into the optical/NIR by non-thermal scattering and fluorescence in order to make up for the reduction in thermally excited flux in the optical/NIR.  However, we find that \citet{fransson15} still significantly underpredicts the flux emitted in the optical/NIR by a factor of $\sim 3$.

Our detection makes it more likely that the flattening observed by \citet{graur16} in the SN~2012cg broadband $F350LP$ light curve  was also due to $^{57}\mathrm{Co}$ rather than a light echo. \citet{graur16} estimated 
$\log (^{57}\mathrm{Co}/^{56}\mathrm{Co}) = -1.36^{+0.11}_{-0.13} $
for SN~2012cg, which is in mild disagreement with our estimate for SN~2011fe.  However, SN~2012cg was only observed in the optical with a very broad long-pass filter.  Limiting our analysis to only our optical observations of SN~2011fe, we find $\log (^{57}\mathrm{Co}/^{56}\mathrm{Co}) = \CotoNifitopt$.  This is in better agreement, but there is a $\sim 1\sigma$ disagreement.  It is not clear if this difference is real, if it is due to a changing SED of SN~2012cg, if it is the result of the metallicities of the progenitors of SN~2011fe and SN~2012cg, or if it is caused by some other intrinsic difference between these two SNe.  

Additional theoretical work is needed to determine the possible isotopic yields and predicted late-time light curves from a broader range of progenitor systems.  Numerous questions remain. (1) How do the masses and the mass ratios of the merging WDs affect the abundance yields?  \citet{roepke12} only modeled the merger of a $1.1$ and a $0.9$ \msun{} WDs, but clearly other mergers are possible. (2) Do colliding WDs produce different abundance ratios than the violent merger modeled by \citet{roepke12}?  (3) What affects to asymmetric ejecta geometry and clumping in the ejecta have on the late-time light curve?  (4) Is the assumption of the instantaneous deposition of energy still valid?   (5) What effect does the progenitor's metallicity have on the predicted abundances?  \citet{roepke12} used electron fractions and initial compositions corresponding to solar metallicity for the SD and DD models, respectively, but SN~2011fe was likely subsolar. Using the \citet{bresolin07} oxygen gradient for M101, \citet{stoll11ATEL} determined the gas-phase metallicity at the position of SN~2011fe to be 12 + log(O/H) $= 8.45 \pm 0.05$, or approximately 0.39 or 0.58 of solar using the solar oxygen abundance from \citet{delahaye06} or \citet{asplund09}, respectively.  Additionally, \citet{mazzali14} required subsolar metallicities to model the outermost layer of the ejecta and \citet{graham15a} also found evidence for a subsolar metallicity through a comparative analysis of photospheric and nebular-phase spectra of SN~2011fe and SN~2011by.  Clearly comprehensive theoretical studies on the effects of metallicity are warranted. 

Additional observations of SN~2011fe by HST are also crucial.  Continued observations will further constrain the $^{55}\mathrm{Fe}$-to-$^{57}\mathrm{Co}$ abundance ratio, possibly leading to the detection of a $^{55}\mathrm{Fe}$-powered SNe Ia light curve for the first time.  Additionally, as SN~2011fe fades, a shock-heated non-degenerate companion may be revealed. 
How long will SN~2011fe be visible?  To calculate a rough lower bound, we first estimated the HST WFC3 UVIS limiting magnitudes for $F438W$, $F555W$, and $F600LP$ using the WFC3 UVIS Imaging exposure time calculator (ETC\footnote{\url{http://etc.stsci.edu/etc/input/wfc3uvis/imaging/}}).  We assumed the \citet{taubenberger15} $1016$-day MODS/LBT spectrum of SN~2011fe for the input SED, our measured location of SN~2011fe, an observing date of 2018 July 1, and 5 orbits of observations with 53 minutes of observing time per orbit.  We found that the magnitudes corresponding to a signal-to-noise ratio of 3 for the $F438W$, $F555W$, and $F600LPW$ filters are $\sim \Blimit{}$,  $\sim \Vlimit{}$, and $\sim \RIlimit{}$ mag, respectively.  We then made the extremely conservative assumption that the light curve of SN~2011fe will continue to fade at the $^{57}\mathrm{Co}$ decay rate with no contribution from $^{55}\mathrm{Fe}$ that would slow down the rate of decay. Under these assumptions, SN~2011fe will remain visible to HST WFC3 observations in $F438W$, $F555W$, and $F600LPW$ for $\sim \Blimitdate{}$,  $\sim \Vlimitdate{}$, and $\sim \RIlimitdate{}$ days after $t_{B {\rm max}}$ or roughly through 2018 May.  This estimate ignores any effects that redistribute the bolometric flux out of the optical. Additionally, SN~2011fe will become fainter than the pre-explosion images of M101, so contamination from host-galaxy light or unrelated sources may limit how long SN~2011fe is observable.  However, we again emphasize that SNe Ia are predicted to produce $0.004$ -- $0.013$ \msun{} of $^{55}\mathrm{Fe}$, which has a half life roughly twice that of $^{57}\mathrm{Co}$ and is expected to become the dominant power source $\sim 1600$ -- $2200$ days after $t_{B {\rm max}}$ \citep{roepke12}.  Thus, SN~2011fe is likely to be observable for longer than our estimate. We will be obtaining two additional epochs of observations for SN~2011fe in  HST Cycles 24, and 25.

While SN~2011fe is likely to remain a uniquely well-studied event, more SN Ia with late-time measurements are needed to constrain the $^{57}\mathrm{Co}/^{56}\mathrm{Co}$ ratio. These measurements are only possible for the nearest and brightest objects.   While such nearby SNe Ia are rare, we are maximizing our probability of discovering them with the All-Sky Automated Survey for Supernovae (ASAS-SN; \citealp{shappee14}).  ASAS-SN monitors the entire visible sky in both the Northern and Southern hemispheres to find nearby SNe with an as minimally biased search as possible. In the upcoming HST Cycle 24, time was also allocated (PI: O. Graur) to add  ASASSN-14lp \citep{shappee16a} and SN~2015F \citep{monard15CBET} to the sample of SNe with multiband observations out to 900--1000 days after $t_{B {\rm max}}$.

\acknowledgments

The authors thank Jill Gerke, Anthony Piro, Or Graur, Armin Rest, Scott Adams, Chris Burns, Ryan Foley, and Jennifer van Saders for discussions and encouragement. We also thank Crystal Mannfolk, Susana Deustua, and the entire space telescope team for their help planning and executing these observations. Finally, we would like to thank our anonymous referee for their careful reading.

This work is supported through HST-GO-14346 and HST-GO-14166. B.S. is supported by NASA through Hubble Fellowship grant HF-51348.001 awarded by the Space Telescope Science Institute, which is operated by the Association of Universities for Research in Astronomy, Inc., for NASA, under contract NAS 5-26555. CSK and KZS are supported by NSF grants AST-1515876 and AST-1515927. 

The LBT is an international collaboration among institutions in the United States, Italy and Germany. LBT Corporation partners are: The Ohio State University, and The Research Corporation, on behalf of The University of Notre Dame, University of Minnesota and University of Virginia; The University of Arizona on behalf of the Arizona university system; Istituto Nazionale di Astrofisica, Italy; LBT Beteiligungsgesellschaft, Germany, representing the Max Planck Society, the Astrophysical Institute Potsdam, and Heidelberg University.  

IRAF is distributed by the National Optical Astronomy Observatory, which is operated by the Association of Universities for Research in Astronomy (AURA) under a cooperative agreement with the National Science Foundation.

Funding for the SDSS and SDSS-II has been provided by the Alfred P. Sloan Foundation, the Participating Institutions, the National Science Foundation, the U.S. Department of Energy, the National Aeronautics and Space Administration, the Japanese Monbukagakusho, the Max Planck Society, and the Higher Education Funding Council for England. The SDSS web site is http://www.sdss.org/.  

This research has made use of NASA's Astrophysics Data System Bibliographic Services.


\bibliographystyle{apj}

\end{document}